\documentclass[apj]{emulateapj}

\begin{document}
\title{The $\beta$-Model Problem: The Incompatibility of X-ray and Sunyaev-Zeldovich Effect Model Fitting for Galaxy Clusters}
\author{Eric J. Hallman}
\author{Jack O. Burns}
\affil{Center for Astrophysics and Space Astronomy, Department of
  Astrophysical and Planetary Sciences, 
University of Colorado, Boulder, CO 80309}
\author{Patrick M. Motl}
\affil{Department of Physics and Astronomy,
Louisiana State University, Baton Rouge, LA 70803}
\author{Michael L. Norman}
\affil{Center for Astrophysics and Space Sciences, University of
California-San Diego, 9500 Gilman Drive, La Jolla, CA 92093}
\email{hallman@casa.colorado.edu}
\slugcomment{Accepted to the Astrophysical Journal}
\begin{abstract}
We have analyzed a large sample of numerically simulated
clusters to demonstrate the adverse effects resulting from use of X-ray fitted $\beta$-model
parameters with Sunyaev-Zeldovich effect (SZE) data. There is a
fundamental incompatibility between $\beta$ fits to X-ray surface
brightness profiles and those done with SZE profiles. Since
observational SZE radial profiles are in short supply, the X-ray
parameters are often used in SZE analysis. We show that this leads to
biased estimates of the integrated Compton y-parameter inside
$r_{500}$ calculated from
clusters. We suggest a simple correction of the method, using a
non-isothermal $\beta$-model modified by a universal temperature
profile, which brings these calculated quantities into closer
agreement with the true values.   
\end{abstract}
\section{Introduction}
The hot gas in clusters of galaxies is responsible for inverse Compton
scattering cosmic microwave background (CMB) photons as they travel through the
intracluster medium (ICM). This results in a spectral distortion of
the CMB at the location of clusters on the sky, referred to as the \citet{sz} effect (SZE). This distortion is characterized by a low
frequency ($<$218 GHz) decrement, and higher frequency ($>$218 GHz) increment in the CMB
intensity \citep{carl}. The X-ray emission in clusters consists of thermal
bremsstrahlung and line emission from the same highly ionized plasma that scatters
the CMB. 

High resolution X-ray or SZE observations of clusters coupled with assumptions
about the gas distribution lead to estimates of the gas mass in the
cluster dark matter potential well. The electron number density is
often assumed to fit a $\beta$ model \citep{caval},
\begin{equation}
  n_{e}(r) = n_{e0}\left(1 + \left(\frac{r}{r_{c}}\right)^{2}\right)^{-3\beta/2}.
\end{equation}
In the above relation, $n_{e0}$ is the central density normalization
and $r_c$ indicates the fitted parameter referred to as the core radius.
Fitting an observed X-ray or SZE profile to these projected $\beta
$ model X-ray surface brightness and SZE $y$ parameter distributions
results in a description of the density distribution, which can be
integrated to obtain the gas mass. The difference in dependence on gas
density and temperature of X-ray emissivity and the SZE $y$ parameter
makes the combination of these two methods of observation potentially very
powerful. Because of this difference, the observability of clusters
via each method is affected differently by the impact of physics in cluster cores including
radiative cooling and feedback mechanisms, as well as the transient
boosting of surface brightness and spectral temperature generated
during merging events \citep{roettmass,motl_sf}. These two methods not only select a
different sample of clusters, but combined SZE/X-ray observations of
individual clusters allow one to extract the density and temperature
of the gas without relying on X-ray spectral temperatures.
\subsection{Isothermal Beta Models}
Under the assumption that the gas in clusters is isothermal, one can
fit an isothermal $\beta$-model to the data in order to deduce the
density profile. To generate the projected X-ray surface
brightness profile, we integrate the $\beta$-model density distribution
\begin{equation}
S_X = \frac{1}{4 \pi (1+z)^4} \int n_e(r) n_H(r) \Lambda(T) dl,
\end{equation}
where in the bremsstrahlung limit, $\Lambda(T) \propto T^{1/2}$, and
in a fixed X-ray band is more weakly dependent on temperature \citep{mohr99}.
This integration results in
\begin{equation}
S_X(b) = S_{X0} \left(1 + \left(\frac{b}{r_c}\right)^2 \right) ^
{\frac{1}{2} - 3\beta}, 
\end{equation}
where $S_{X0}$ is the fitted central X-ray surface brightness
of the model, and $b$ indicates the projected radius.
Similarly for the SZE, a $\beta$-model density profile can be integrated 
\begin{equation}
y = \int \sigma_T n_e(r) \frac{k_b T}{m_e c^2} dl,
\end{equation}
which results in a
projected radial distribution of the Compton $y$ parameter 
\begin{equation}
y(b) = y_0 \left(1 + \left(\frac{b}{r_c}\right)^2\right)^{\frac{1}{2}
  - \frac{3\beta}{2}}.
\end{equation}
For very hot clusters (T $>$ 10 keV), relativistic corrections must be included to
the SZE integral \citep{itoh}.
While these fits result in a description of the cluster density
profile, we must be aware that that description is only approximate,
since it has been shown both in simulations and observations \citep{loken,utp_obs} that
many clusters show a radial dependence of temperature. That means that
when using isothermal models (the above equations) to fit the data, we should expect error
to be introduced in the derived quantities.
\subsection{SZE/X-ray Derived Quantities}
Recent studies have used the values of the
$\beta$-model parameters determined from the X-ray surface brightness
profiles combined with SZE cluster observations to determine the value of the Hubble constant ($H_0$) \citep{reese,bonamente} and the cluster gas fraction \citep{joy,laroque}. While joint fits
of X-ray and SZE interferometric data are used to determine the $\beta$-model
parameters in these studies, the X-ray data drives the fit, since the
SZE data currently lacks the resolution (and interferometric U-V plane
coverage) to constrain the parameters well \citep{laroque}.  

An additional calculation can be performed using combined SZE/X-ray
data from clusters. It is expected both from analytic arguments and numerical simulations
that the integrated SZE signal, as a measure of the total ICM pressure in clusters, should be an excellent
proxy for cluster total mass \citep{dasilva,motl05,nagai,kravtsov06}. This
presupposes that one can determine accurately the value of the
integrated SZE signal to some mass-scaled cluster radius, as well as
perform an accurate calibration of this relationship
observationally. In \citet{motl05}, we showed that the value of
$Y_{500}$, the integrated Compton $y$ parameter inside a radius where
$\delta \rho/\rho$ = 500 (with respect to the critical density) accurately measures the cluster total mass
inside that same radius. In order to accurately calibrate this
relationship, one must measure the $Y$-$M$ relationship to high
precision at low redshift, where clusters have combined SZE/X-ray
observations to use. Then the measured relation, scaled for
redshift, can be used to determine masses of SZE-selected clusters,
which should be indentifiable to high redshifts, in order to constrain
cosmology. 

Since SZE radial profiles are of relatively poor
quality so far, a
direct determination of the compatibility of the X-ray $\beta$-model
parameters with the SZE profiles in clusters can not currently be done
observationally. Here we compare the values of $\beta$-model
parameters for a large number of simulated clusters when fitting the
X-ray surface brightness profile, the SZE radial Compton $y$ profile,
and jointly fitting both X-ray and SZE profiles to a common
$\beta$-model. We show that the use of X-ray parameter values leads to
biased estimates of $Y_{500}$, as well as the integrated gas mass
$M_{500,gas}$. 

We discuss our numerical simulations in Section 2, results of the
analysis in Section 3, the consequences of the use of X-ray
$\beta$-model parameters for estimating the $Y_{500}$-$M_{500}$
relationship for clusters in Section 4, and discussion and conclusions
in Section 5.
\section{Numerical Simulations}
Our simulations use the hybrid Eulerian adaptive mesh refinement hydro/N-body code 
\textit{Enzo} (\citet{enzo}; http://cosmos.ucsd.edu/enzo)
to evolve both the dark matter and baryonic fluid in the clusters,
utilizing the piecewise parabolic method (PPM) for the
hydrodynamics. With up to seven levels of refinement
in high density regions, we attain spatial resolution up to
$\sim \; 16 \; h^{-1}$ kpc in the clusters. We assume a concordance
$\Lambda$CDM cosmological model with the following parameters:
$\Omega_{\mathrm{b}} = 0.026$, $\Omega_{\mathrm{m}} = 0.3$,
$\Omega_{\Lambda} = 0.7$, $\mathrm{h} = 0.7$, and $\sigma_{8} = 0.9$.
Refinement of high density regions is performed as described in \citet{motl04}. 

We have constructed a catalog of AMR refined clusters identified in the simulation
volume as described in \citet{loken}. The catalog of clusters used in
this study includes the effects of radiative cooling, models the loss of low
entropy gas to stars, adds a moderate
amount of supernova feedback due to Type II supernovae in the zones
where stars form, and is identified as the SFF (Star Formation with Feedback) catalog in \citet{hall06}. The catalog includes clusters with total mass
(baryons + dark matter) greater than $10^{14} M_{\odot}$ out to z=2 in
the simulation. This catalog
includes roughly 100 such clusters at z=0, and has 20 redshift
intervals of output, corresponding to a total of roughly 1500 clusters
in all redshift bins combined. 

For parts of the analysis, we have cleaned the cluster sample
as described in \citet{hall06}, removing cool core clusters and
obviously disturbed clusters.  This is done primarily to more closely mimic
observational studies of cluster properties. We examine by eye all cluster projections and
remove any with obvious double peaks in the X-ray or SZE
surface brightness images, have disturbed morphology within R =
$1h^{-1}$ Mpc, exhibit edges consistent with shocked gas, or those
that have cool cores.  The cool core clusters are identified as those
with a projected emission-weighted temperature profile which declines at small
radius, or those with strongly peaked X-ray emission. The cool core
clusters are eliminated because as we have shown in \citet{hall06},
they lead to strong biases and increased scatter in estimates of
cluster physical properties when standard observational methods are
used. We show in our previous work that eliminating obviously
disturbed clusters has a minor, but measurable effect on
observationally derived quantities. Indeed the simple assumptions
that are typically made in the observational derivation (e.g., spherical
symmetry, hydrostatic equilibrium) clearly do not apply
to such clusters. We also use two orthogonal
projected images for each cluster. The final cleaned sample contains 493
cluster projected images from a series of evolutionary epochs from z=0
to z=2.  
\section{Results}
\subsection{Isothermal $\beta$-model Fits}
\begin{figure*}
\begin{center}
\includegraphics[width=0.9\textwidth]{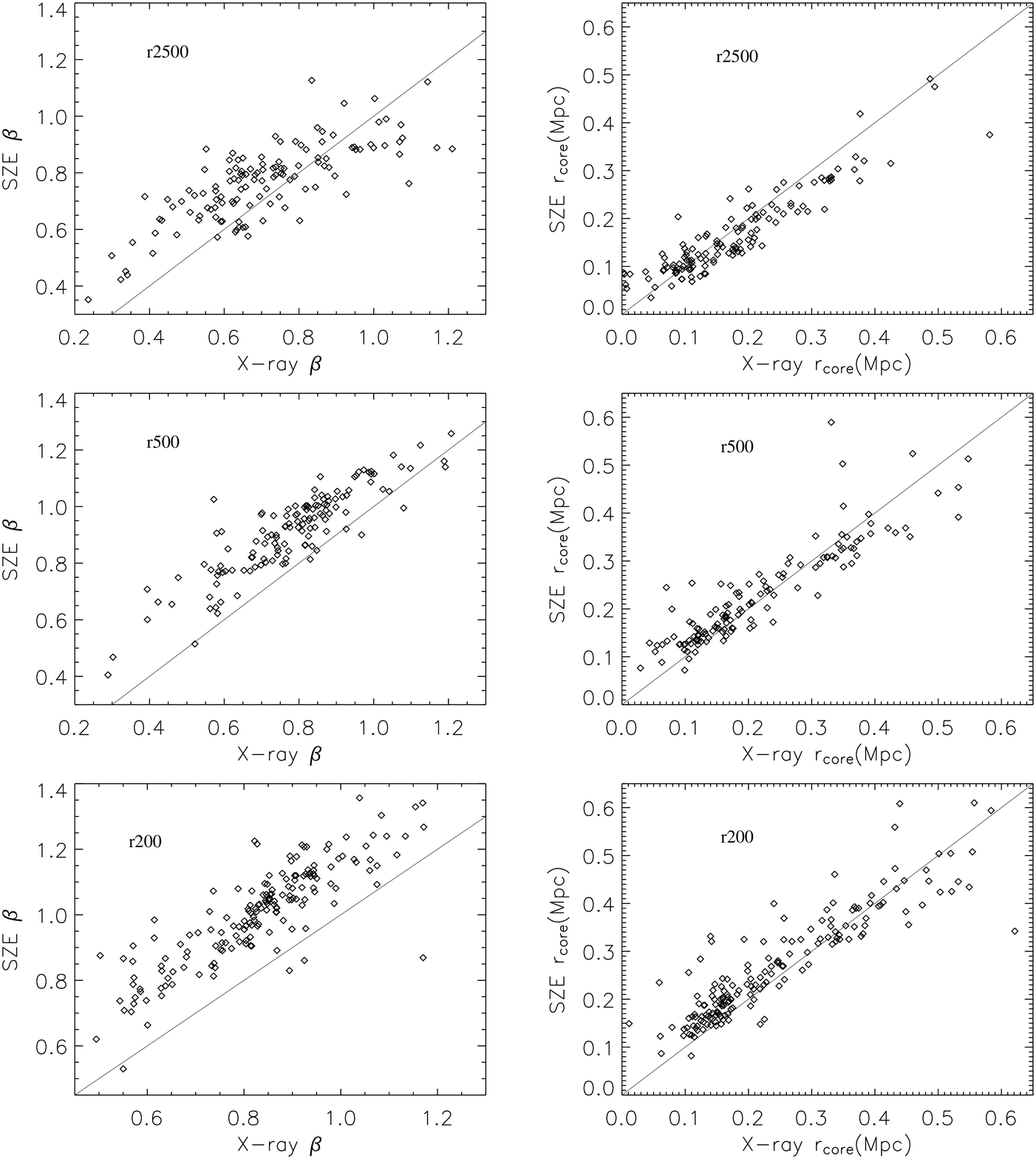}
\vspace{-5mm}
\end{center}
\caption{Upper panels: Left panel shows comparison of fitted $\beta$
values for clusters in our numerical sample at z=0 out to $r_{2500}$ using
the X-ray and SZE images. Right panel shows comparison of fitted value
of $r_{c}$ for same clusters.  Middle panels: Same as above, except
values are for profiles fitted out to $r{500}$. Lower panels: Same as
above, but for profiles fitted out to $r_{200}$.}
\label{6panel}
\end{figure*}
We have generated images of our simulated cluster catalog by
projecting the physical quantities on the grid to get the value of the
SZE Compton $y$ parameter and the X-ray surface brightness. For the
X-ray, we have used a simple bremsstrahlung emissivity. It is a
simpler calculation than using a model X-ray emissivity, and we will
show later that the values of the $\beta$-model parameters are nearly
identical irrespective of which emissivity calculation we use. 

From these images, we have created radial profiles in annular bins,
which we subsequently fit to $\beta$-model profiles. Each
cluster, for both the SZE and X-ray profile, has a set of
$\beta$-model parameters that describe it, though there is some
degeneracy in the parameters in each case. The three $\beta$-model
parameters, namely the normalization ($S_{X0}$ or $y_0$), the value of
the core radius ($r_{core}$), and the power law index $\beta$ are left
as free parameters in the fit.  We have fit to each of three limiting
outer radii in both the SZE and X-ray case, $r_{2500}$, $r_{500}$, and
$r_{200}$. In each case, the subscript indicates the average
overdensity with respect to critical inside that radius. This radius
is calculated from the simulation data using the overdensity of the
dark matter. We have used the full sample for this part of the analysis,
but have excluded the cool core region (typically $\approx$100kpc) of the profiles from the
fitting procedure.  

We have found that there are significant differences in the values of
the fitted model parameters depending on whether the X-ray or SZE
profiles are used. Figure \ref{6panel} shows
the comparison of the values of $\beta$ and $r_{core}$ plotted against
one another for individual z=0 clusters in our catalog for the three
limiting outer radii used. It is clear that there is a large amount of
scatter in the relationship for the $\beta$ values, and also at
$r_{500}$ and $r_{200}$, a definite discrepancy. Fitting SZE profiles results in a
  consistently higher value of $\beta$ than does fitting X-ray
  profiles. While there is scatter in the compared values of
$r_{core}$, there is general agreement within the errors.
  However, since there is a degeneracy between $r_{core}$
  and $\beta$ in the fitting, some scatter is expected. Since the
values of $\beta$ and $r_{core}$ are used directly in the equation for the
density profile, inconsistent values for these parameters lead to
different deduced density profiles, and discrepant values
for the cluster gas mass. 

\begin{figure}
\begin{center}
\includegraphics[width=0.5\textwidth]{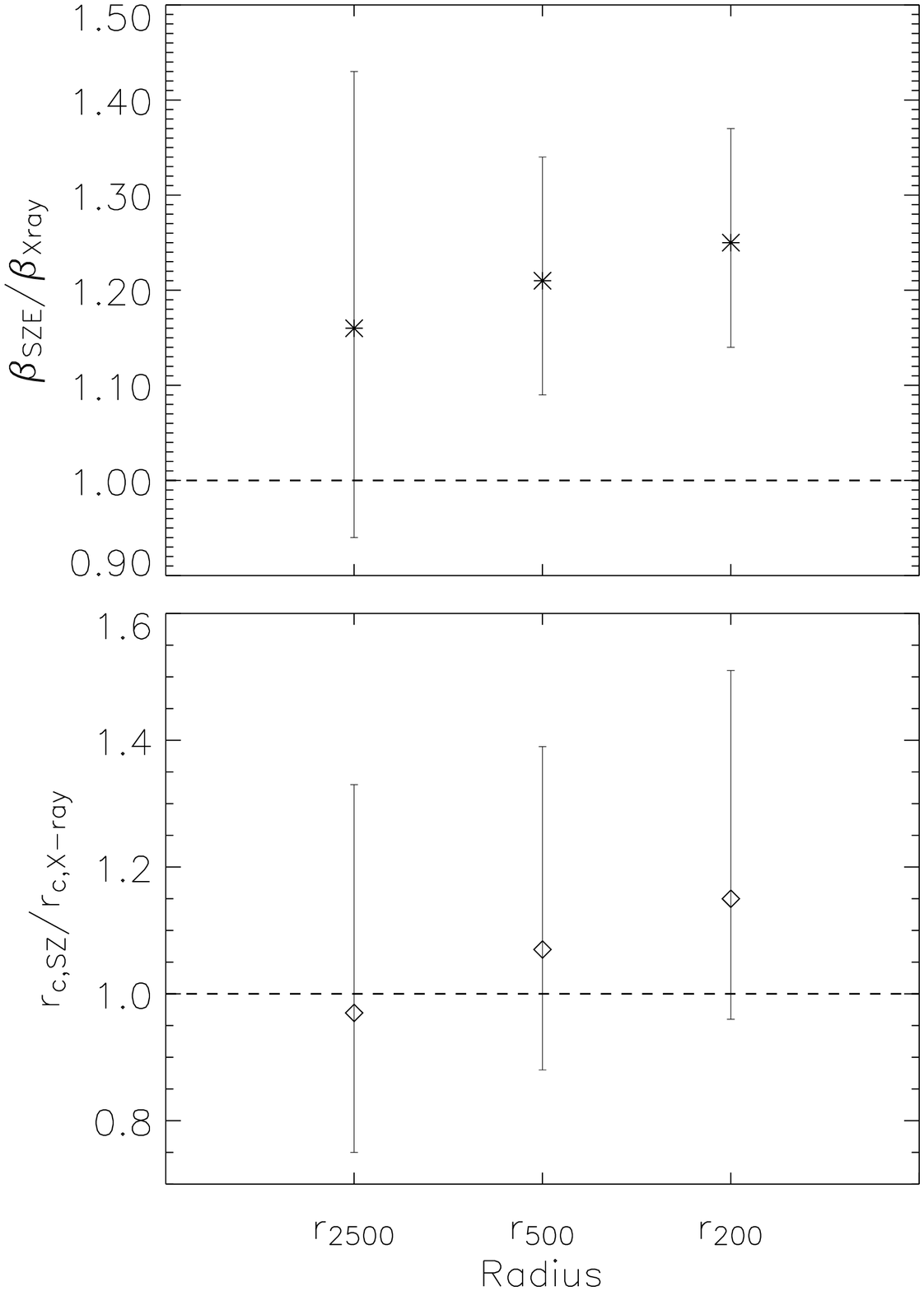}
\end{center}
\vspace{-5mm}
\caption{Upper: Ratio of SZE $\beta$ fitted to
radial profiles from projections of full SFF simulated sample of
clusters to $\beta$ fitted to X-ray radial profiles for the same
clusters. Median values and 1$\sigma$ error bars for the ratio
$\beta_{SZE}/\beta_{X-ray}$ in z=0 clusters fitted to each of three radii $r_{2500}$, $r_{500}$,
and $r_{200}$ using the isothermal $\beta$-model. Lower: Median values and 1$\sigma$ error bars for the
ratio $r_{core,SZE}/r_{core,X-ray}$ in z=0 clusters for the three radii $r_{2500}$, $r_{500}$,
and $r_{200}$.}
\label{comp_rad}
\end{figure}
Next, looking at Figure \ref{comp_rad}, we show the 1$\sigma$ scatter
in the ratio of the SZE deduced parameters and the X-ray parameters
for each of our three fiducial radii for all clusters in our catalog. While there is not a
statistically significant bias in $r_{core}$, one clearly exists in $\beta$ at the larger radii. To answer the question of why the isothermal $\beta$-model fit for
X-ray and SZE observations give a different set of model parameters in
each case, consider the simple argument in the next section.
\subsection{The Problem with the Isothermal Model}
The X-ray and SZE surface brightness functions resulting from the isothermal
$\beta$-model density profile integration have distributions
at $r>r_c$ which depend on radius as
\begin{equation}
S_X(r) \propto r^{1-6\beta}
\end{equation}
and
\begin{equation}
y(r) \propto r^{1-3\beta}.
\end{equation}

 We have fit the
temperature profile of our simulated clusters to a universal
temperature profile (UTP) of the form
\begin{equation}
  T(r) = \langle T \rangle_{500}T_{0}\left(1 + \left(\frac{r}{\alpha r_{500}}\right)^{2}\right)^{-\delta},
\end{equation}
where $T_{500}$ indicates the average spectral
(in this study, emission-weighted) temperature inside $r_{500}$. $T_0$, $\alpha$, and
$\delta$ are dimensionless fitted parameters to the spherically averaged (from the
three-dimensional simulated data) temperature profiles of all clusters at
each redshift in the simulations used in this study. The mean values
of these parameters for clusters at select redshifts from our
simulations are shown in Table
\ref{par_tab}. We have used the redshift-specific mean value for
each cluster in the analysis.
\begin{table}
\caption{Mean parameters for Universal Temperature Profile}
\begin{tabular}{cccccc}
\hline
\hline
Parameter & $z=0$ & $z=0.1$ & $z=0.25$ & $z=0.5$ & $z=1.0$ \\
\hline
$T_0$ & 1.25$\pm$0.06\tablenotemark{a} & 1.27$\pm$0.06 & 1.31$\pm$0.08 & 1.37$\pm$0.09 &
1.37$\pm$0.14 \\
$\delta$ & 0.51$\pm$0.21 & 0.51$\pm$0.19& 0.42$\pm$0.13& 0.43$\pm$0.12
& 0.53$\pm$0.22\\
$\alpha$ & 1.17$\pm$0.46 & 1.12$\pm$0.40& 0.91$\pm$0.34& 0.80$\pm$0.28
& 0.90$\pm$0.45 \\
\hline
\end{tabular}
\label{par_tab}
\tablenotetext{a}{Error bars indicate 1$\sigma$ dispersion.}
\end{table}

If the cluster's true gas temperature declines with radius with the
above described dependence
\begin{equation}
T(r) \propto r^{-2\delta},
\end{equation}
then the cluster observable profiles have dependence at $r>r_c$ 
\begin{equation}
S_X(r) \propto r^{1-6\beta-\delta}
\end{equation}
and 
\begin{equation}
y(r) \propto r^{1-3\beta-2\delta}.
\end{equation}
Our simulations show a typical value of $\delta=0.5$ (see Table \ref{par_tab}). 
If we assume the true cluster density profile is a $\beta$-model
modified in this way by a UTP for the temperature profile (and
$\delta$=0.5), and that the true value for $\beta$ set by the cluster
density profile is $\beta$=0.8, then the radial dependence of X-ray
surface brightness and SZE surface brightness are
\begin{equation}
S_X(r) \propto r^{-4.3}
\end{equation}
and 
\begin{equation}
y(r) \propto r^{-2.4}.
\end{equation}

Finally, if we then fit an isothermal $\beta$-model to cluster
profiles with the above dependence, setting powers equal for the X-ray 
\begin{equation}
1-6\beta = -4.3
\end{equation}
we would get $\beta = 0.88$, 
and for the SZE, the isothermal fit would give us
\begin{equation}
1-3\beta = -2.4
\end{equation}
or $\beta = 1.13$.

While the declining temperature profile has a relatively
small effect on the $\beta$ value extracted from an isothermal fit to
the X-ray surface brightness (+10\%), there is a larger effect on the
value of $\beta$ in the SZE case (+41\%). This is not surprising given
the difference in temperature dependence between Equations 2 and
4. Table \ref{betas} shows the
median values of fitted $\beta$ in our cluster sample out to
$r_{200}$, with 1$\sigma$ scatter. Indeed the variation between the
X-ray and SZE fits is consistent with the simple analysis shown
above. Also, one line of the table shows the result of fitting
the true density profiles extracted from the simulated data. Though
the differences between the true value and the SZE and X-ray fitted
values are bigger than expected from a simple analysis, the true value
is indeed smaller than that fitted from the emission profiles. There
are additional sources of bias which may be introduced due to
clumpiness in the ICM and deviations from spherical symmetry which may
contribute to the larger difference in fitted parameters
\citep{sulkanen, nagai00}. There is
also the degeneracy in $r_c$ and $\beta$ in any model fitting to these
profiles which can contribute to differences. We also show in the last
line of the table the result of fitting the isothermal $\beta$-model
to the profile of X-ray surface brightness resulting from a projection
of the Raymond-Smith \citep{rs} model emissivity. It is clear that the
statistical properties of the fitted parameters are nearly
indistinguishable from those fitted to the simple bremsstrahlung model.

It is therefore clear that the effect of a declining temperature
profile in real clusters is to alter the fitted values of
$\beta$ in this simple model, such that the SZE isothermal $\beta$ is larger
than both the X-ray, and the ``true'' value associated with the
density profile of the cluster.  Indeed a similar analysis by
\citet{ameglio} has shown results consistent with ours in this regard.  
We also find no significant trend in the SZE/X-ray parameter ratios
with redshift.   
\begin{table}
\caption{Median Parameter Values from Profile Fitting to $r_{200}$}
\begin{center}
\begin{tabular}{ccccccc}
Method & $\beta$ & +1$\sigma$ & -1$\sigma$ & $r_{core}$(kpc) & +1$\sigma$ & -1$\sigma$\\
\hline
Iso X-ray & 0.84 & 1.02 & 0.70 & 168 & 330 & 99\\
Iso SZE & 1.05 & 1.27 & 0.88 & 196 & 340 & 130\\
U-$\beta$ X-ray & 0.81 & 0.97 & 0.67 & 165 & 320 & 96\\ 
U-$\beta$ SZE & 0.82 & 0.97 & 0.69 & 160 & 260 & 103 \\
Density & 0.70 & 0.87 & 0.55 & 137 & 242 & 73 \\ 
Iso Raymond-Smith & 0.82 & 1.03 & 0.67 & 168 & 318 & 96 \\
\hline
\end{tabular}
\label{betas}
\end{center}
\end{table}
\subsection{Fitting to a UTP Modified Model}
Since the variation in model parameters between X-ray and SZE fitting
appears to be due to the failure to account for the radial dependence of
temperature in the intracluster medium, it makes sense to use a model
that includes this radial dependence for fitting the surface
brightness. Our non-isothermal $\beta$-model for the surface
brightness is created by integrating the expression for the X-ray
surface brightness and Compton $y$ parameter using the standard
$\beta$-model for the density, and the UTP for the
temperature. We refer to this model as the U-$\beta$. Integrating Equations 2 and 4 with the substitution of
the $\beta$-model density profile (Eq. 1) and the UTP (Eq. 8) results
in the fitting relations for the X-ray and SZE surface brightness
\begin{equation}
S_X(b) = S_{X0} \left( 1 + \left(\frac{b}{r_c} \right)^2
\right)^{-3\beta} \left(1+\left(\frac{b}{\alpha r_{500}} \right)^2
\right)^{-\frac{\delta}{2}} I_X(b)
\end{equation}
and
\begin{equation}
y(b) = y_{0} \left( 1 + \left(\frac{b}{r_c}\right)^2
\right)^{-\frac{3\beta}{2}} \left(1+\left(\frac{b}{\alpha r_{500}} \right)^2
\right)^{-\delta} I_{SZ}(b),
\end{equation} 
respectively, where $I_X(b)$ and $I_{SZ}(b)$ are line integrals described in the Appendix.

\begin{figure}
\begin{center}
\includegraphics[width=0.5\textwidth]{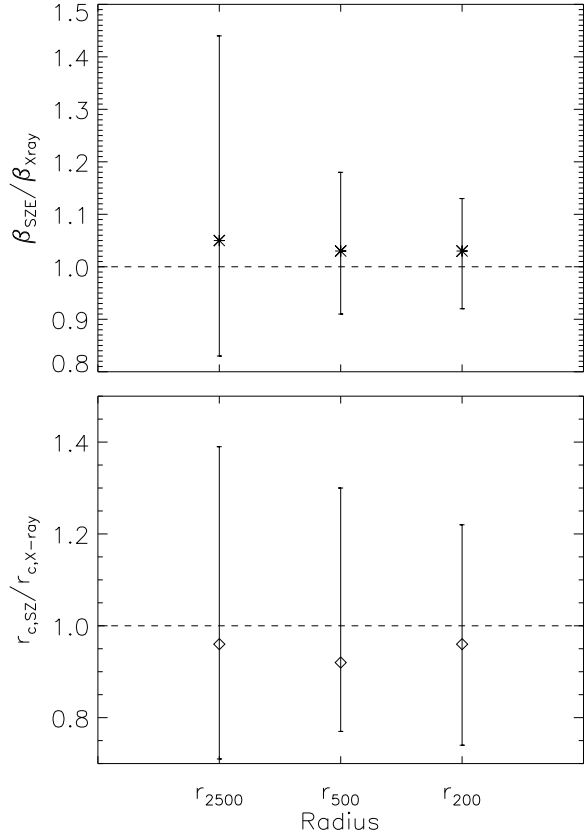}
\vspace{-5mm}
\end{center}
\caption{Upper: Ratio of SZE $\beta$ fitted to
radial profiles from projections of full SFF simulated sample of
clusters to $\beta$ fitted to X-ray radial profiles for the same
clusters. Median values and 1$\sigma$ error bars for the ratio
$\beta_{SZE}/\beta_{X-ray}$ in z=0 clusters when fitted to each of three radii $r_{2500}$, $r_{500}$,
and $r_{200}$ when fitting with a U-$\beta$ model. Lower: Median values and 1$\sigma$ error bars for the
ratio $r_{core,SZE}/r_{core,X-ray}$ in z=0 clusters for the three radii $r_{2500}$, $r_{500}$,
and $r_{200}$.}
\label{comp_rad_utp}
\end{figure}

As described in Section 3.2, the parameters $\alpha$ and $\delta$ are
fixed in the fitting to the surface brightness distributions, and
result from fitting of the spherically averaged (from the
three-dimensional simulated data) temperature profiles from
the simulated clusters. The average values of the parameters in each
redshift bin are used
for fitting the surface brightness profiles of a cluster in that same
redshift bin. When fitting the cluster profiles to these relations, we find that
there is now consistency in the values of $\beta$ and $r_c$ between
X-ray and SZE fitting, shown in Figure \ref{comp_rad_utp}. While there is still some scatter, the offset is
removed. Table \ref{betas} shows the results for all clusters in the
sample when fitting to an isothermal $\beta$-model or a non-isothermal
(the U-$\beta$) model. The value of $\beta$ is virtually identical in SZE and
X-ray fitting when using the U-$\beta$. Additionally, they result in similar
distributions of $r_c$ values.

Note that this model has the same number of free parameters (three) used to
fit the surface brightness as a standard $\beta$-model, the others are
fixed from simulations. We should note the well-known caveat about
$\beta$-model fitting, that the values of $r_{core}$ and $\beta$ are
somewhat degenerate in this fitting.  
\section{Consequences of the $\beta$ Incompatibility: Example Calculation of $Y_{500}$ vs $M$}
Using the isothermal $\beta$-model parameters from a fit dominated by the X-ray data to do
SZE or combined X-ray/SZE analysis introduces an additional error or
bias to various derived quantities. Here, we characterize
the nature of these errors on the measured $Y_{500}$ vs $M_{500,gas}$
relation for clusters. As described in the introduction, it has been
shown that the integrated Compton $y$ parameter inside $r_{500}$ is an
excellent proxy for total mass
\citep{dasilva,motl05,nagai,kravtsov06}. We examine whether one can accurately
determine the true value of $Y_{500}$ (and $M_{500,gas}$) from a
cluster observation using standard observational techniques.

Generating a value for $Y_{500}$ using a $\beta$-model fit
dominated by the X-ray emission, one must extrapolate to $r_{500}$
using the model for the SZE emission. Currently existing cluster SZE
profiles from interferometric instruments do not constrain individual
$\beta$-model parameters well. Since we have shown that the
isothermal $\beta$-model paramters for SZE and X-ray cluster profiles are
different, there is an error introduced. Additionally, there are
errors introduced in the calculation of the cluster gas mass, $M_{500,gas}$, since in
any individual cluster, the $\beta$-model for the density deduced from
the observations is only approximately correct. 
\subsection{Estimation of Y and M}
We have used the simulated cluster sample described above (SFF) to determine the
systematic errors introduced in this procedure. We have analyzed the
simulated clusters such that our synthetic observations are analogous to
unbiased, high signal-to-noise observations of real clusters. The simulated
observations are idealized, since no instrumental effects
or foreground/background source removal are simulated. We have performed our calculations on projected SZE and
X-ray images generated from each simulated cluster, assuming the
X-ray profiles could be determined to $r_{500}$.
 
Using this method, and assuming that the values of $\beta$ and $r_c$
are defined by fitting of the X-ray surface brightness profile, we have characterized errors by taking the model parameters from
the X-ray fitting and calculated the estimated values of $Y_{500}$
and $M_{500,gas}$ from each cluster. We then compare the value to the true
value taken from the simulation grid. 

We have used the isothermal
model fits and also the fits to the U-$\beta$ models. In
each case, $Y_{500}$ is determined by extrapolating the model for
$y(r)$ using the parameters $\beta$ and $r_c$ from the X-ray fitting,
and the value of $y_0$ generated by fitting the SZE profile with
$\beta$ and $r_c$ fixed. $M_{500,gas}$ is estimated by integrating the
$\beta$ model for density out to an overdensity of 500. This method
effectively measures $r_{500}$ from the fitted gas density profile in order to
get the integrated gas mass. For the
U-$\beta$ cases, the values of $\alpha$ and $\delta$ are fixed at the mean
values for the whole sample of clusters at each redshift, recognizing
that these values would be provided by simulations, and would not be
left as free parameters in the observational analysis. 
\subsection{Comparison of Isothermal and U-$\beta$ Methods}
\begin{figure}
\includegraphics[width=0.5\textwidth]{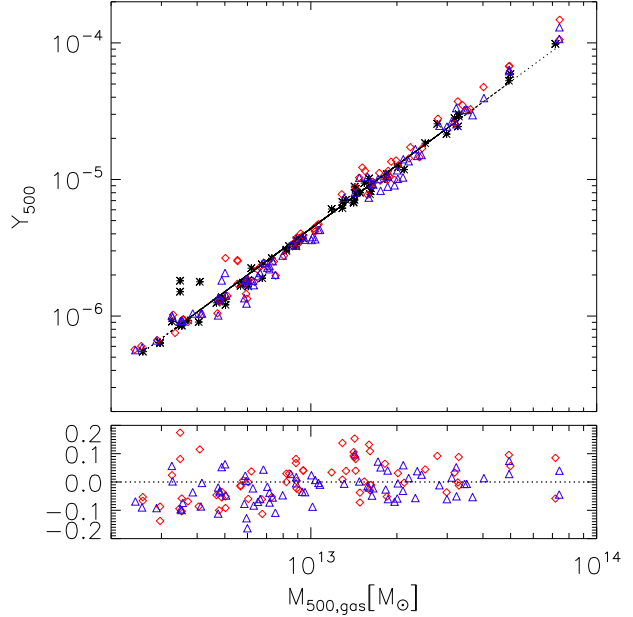}
\caption{Upper: Integrated Compton $y$ parameter inside $r_{500}$ from
  simulated clusters at z=0 epoch plotted against integrated mass inside the same
  radius. Stars are for the true values on the simulation grid,
  red diamonds represent values deduced by fitting to isothermal
  $\beta$-models, blue triangles are for fitting to a U-$\beta$ model. Lower: Log deviation of isothermal and U-$\beta$ points
  from the best-fit scaling relation for the true values.}
\label{yvalues}
\end{figure}
On first inspection of Figure \ref{yvalues}, it is not clear that
either the isothermal or U-$\beta$ generates a more accurate
result. The plot shows values from all clusters from the z=0
simulation epoch. The true values of $Y_{500}$ and $M_{500,gas}$ are plotted as
stars, the isothermal model values as diamonds, and the U-$\beta$ model
values as triangles. 

The plot of deviation of these values from the best fit to the
true values shows some difference between these
two methods. However, it is important to remember that there are
deviations from both the true integrated SZE, and the true gas mass,
so these points deviate in both dimensions of the plot. Since errors
in both are correlated, due to the dependence on the determination of
$r_{500}$ from the $\beta$-model, this should reduce the apparent
separation of points from the best fit, even when the values of both
parameters are in error. 
\begin{figure}
\includegraphics[width=0.5\textwidth]{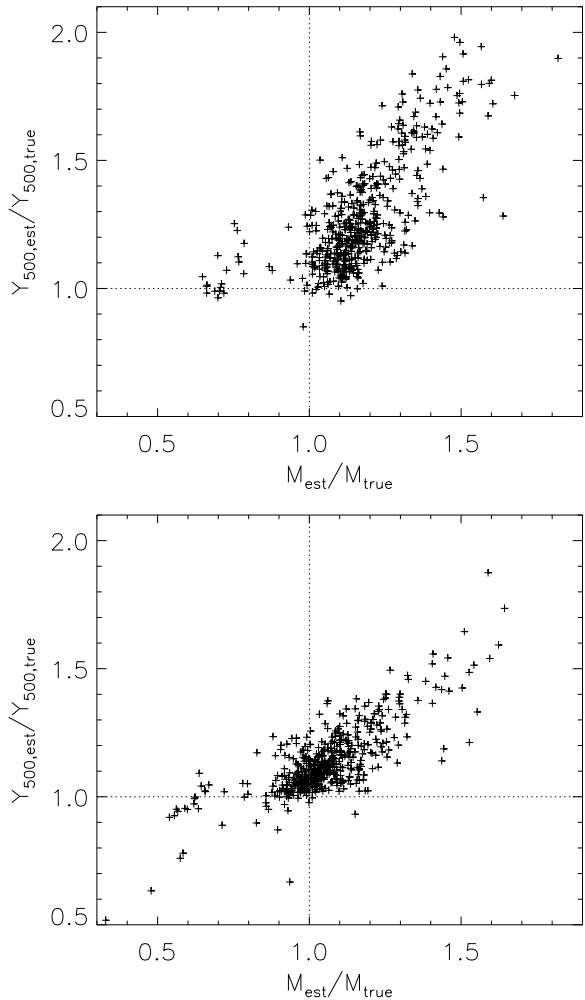}
\caption{\small Upper: Plot of the ratio of estimated $Y_{500}$ to its true value
  (y-axis) versus the ratio of estimated $M_{500,gas}$ to its true value
  for 493 simulated cluster projections for isothermal $\beta$-model fits. 
  Lower: Ratio of estimated $Y_{500}$ to its true value
  versus the ratio of estimated $M_{500,gas}$ to its true value for U-$\beta$-model fits. Lines indicate location of perfectly accurate estimation
  of each quantity.}
\label{yvsm}
\end{figure}

\begin{figure}
\includegraphics[width=0.5\textwidth]{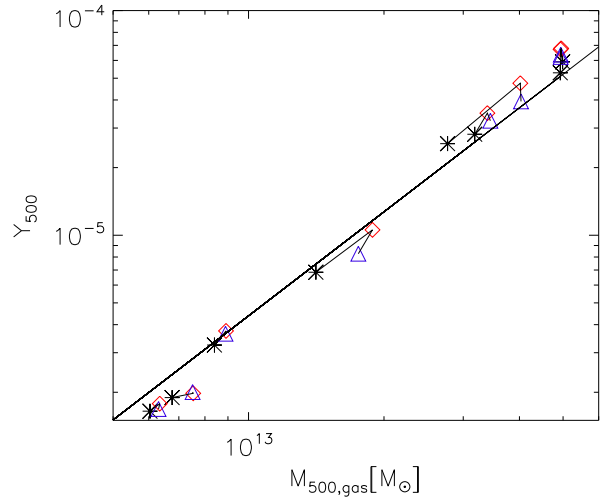}
\caption{Plot of eight clusters on the $Y_{500}$-$M_{500,gas}$
  relation chosen to span a wide mass range, but otherwise randomly. Stars indicate the true values from the simulation, red
  diamonds the isothermal $\beta$-model estimates, and blue triangles
  the U-$\beta$ model estimates. Lines link points
  associated with the same cluster.}
\label{shift}
\end{figure}

This important point is illustrated in Figure \ref{yvsm}, which shows
the the ratios of estimated to true values of $Y_{500}$ and $M_{500,gas}$
for each cluster. In both the isothermal (upper plot) and U-$\beta$
(lower plot) methods, there is a clear trend in over and
underestimation in both values. These figures also illustrate nicely
the improvement in values one gets with the U-$\beta$ method. The U-$\beta$
points are clustered more strongly around the correct values than are
the isothermal estimates. This effect is illustrated in Figure \ref{shift}, where we
have plotted the true values of $y$ and $M$, with lines indicating the
corresponding estimates of those values for the isothermal and U-$\beta$
methods. There are typically overestimates in both estimated $Y_{500}$
and $M_{500,gas}$, and the use of the U-$\beta$ typically brings the estimated
values back toward the true.  

The outliers in these plots are also interesting. The clusters which
give strong overestimates of $Y$ and $M_{gas}$ typically result from
line-of-sight overlap of multiple structures. This enhances the
integrated SZE signal in projection, but overestimates the true value
for the main cluster. The clusters with low estimated values of $Y$
and/or $M_{gas}$ typically have poor quality $\beta$-model fits (as
measured by a $\chi^2$ statistic), and appear to have
systematically low fitted values for the core radius. Some of these
clusters appear to have weak cool cores, not meeting the cool core
criteria, and thus were not excluded from the sample. It is interesting
that the clusters with mass underestimates do not all generate
underestimates of $Y$. This effect warrants further study.

Table \ref{y500_tab} also shows the
result of this analysis. When using isothermal models, the median
estimated value of $Y_{500}$ is 23\% larger than the true value (the
mean is a 28\% overestimate). Additionally, when estimating the gas
mass, the median estimated value is a 16\% overestimate. In previous
work \citep{hall06}, we have shown that the magnitude of the
overestimate via X-ray $\beta$-model methods is lower than this, but
in that case we assumed one could correctly calculate the value of
$r_{500}$ with no error. In either case, this overestimate is
consistent with our previous work, and with that of others
\citep{mohr99, math99}. The overestimate in mass results from
substructure, merging and other physical processes not described by a
simple model. In the current analysis, the value of
$r_{500}$ is deduced directly from the $\beta$-model. In contrast, the
U-$\beta$ method gets closer to the correct value for $Y_{500}$, with a
median value 11\% higher than the true value. For the mass, the U-$\beta$
method gets an estimate close to the true mass, a median overestimate of only 4\%. 

\begin{figure}
\includegraphics[width=0.5\textwidth]{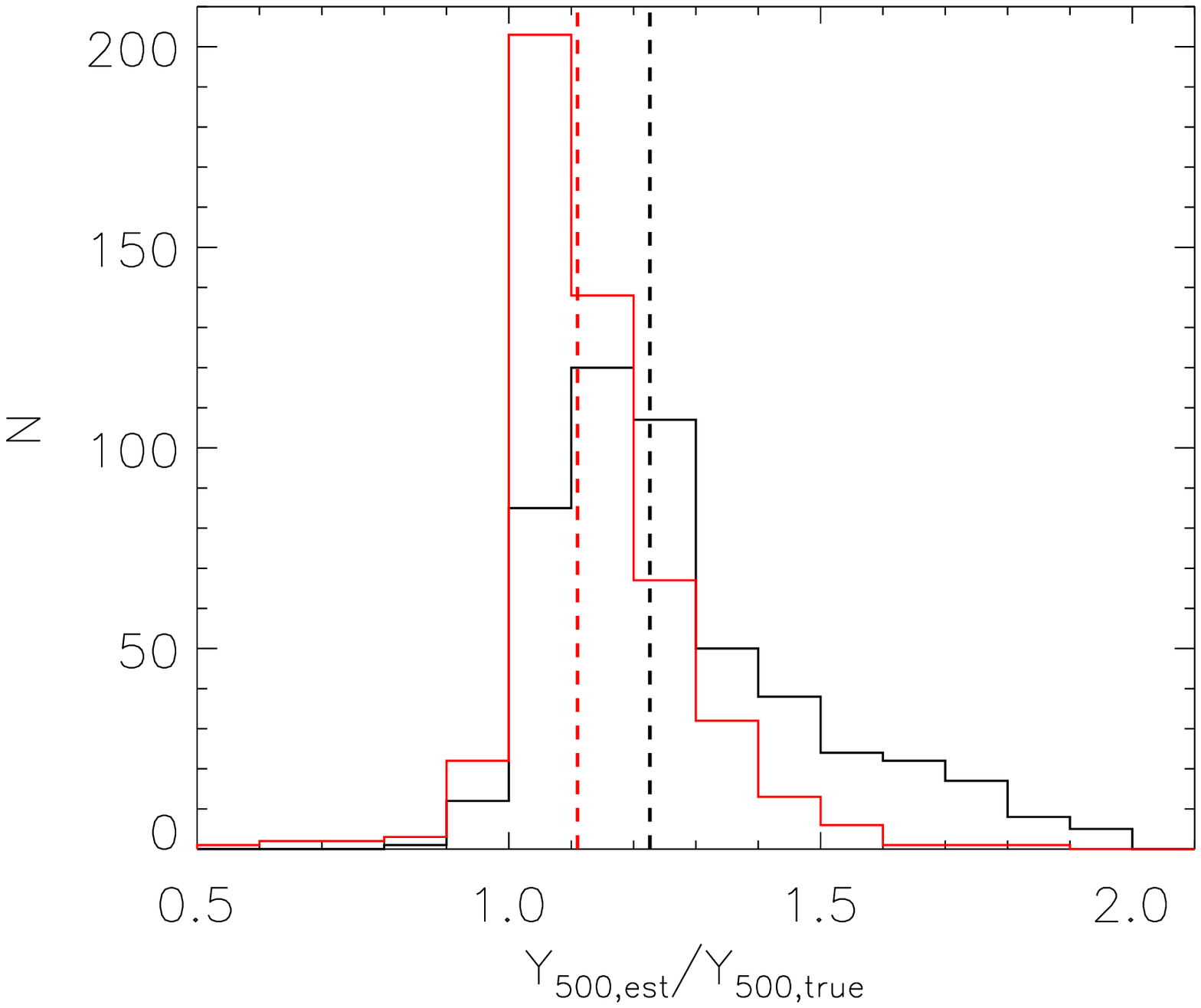}
\caption{Histogram of 493 simulated cluster projections of ratio of estimated
  $Y_{500}$ to true value for isothermal (black lines) and U-$\beta$ (red
  lines) methods. Dotted lines indicate median values for each distribution.}
\label{yhist}
\end{figure}

\begin{figure}
\includegraphics[width=0.5\textwidth]{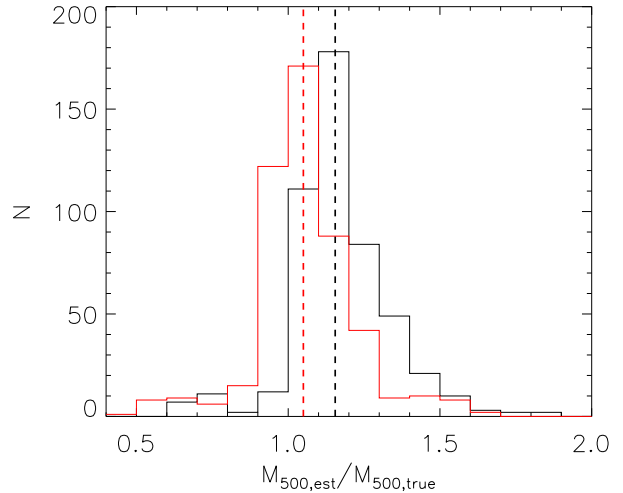}
\caption{Histogram of 493 simulated cluster projections of ratio of estimated
  $M_{500,gas}$ to true value for isothermal (black lines) and U-$\beta$ (red
  lines) methods. Dotted lines indicate median values for each distribution.}
\label{mhist}
\end{figure}

Additionally, using the U-$\beta$ method reduces the scatter in $Y_{500}$
values, as shown in Figure \ref{yhist}. The distribution of values is
more sharply peaked, with a smaller high end tail in addition to a
reduced bias compared to the isothermal method. In mass, the result
is a reduced bias in the median values with a slight improvement in
the scatter as shown in Figure \ref{mhist}.
\begin{table}
\caption{Ratio of Estimated to True Values of Simulated Cluster Properties to $r_{500}$}
\begin{center}
\begin{tabular}{ccc}
 & Isothermal & U-$\beta$ \\
\hline
$\langle y_{500,est}/y_{500,true} \rangle$ & 1.28 & 1.13 \\
$(y_{500,est}/y_{500,true}), median$& 1.23 & 1.11\\
1$\sigma$ upper & 1.50 & 1.24\\
1$\sigma$ lower & 1.08 & 1.03\\
$\langle M_{500,gas,est}/M_{500,gas,true} \rangle$ & 1.17 & 1.05 \\
$(M_{500,gas,est}/M_{500,gas,true}), median$& 1.16 & 1.04\\ 
1$\sigma$ upper & 1.31 & 1.18\\
1$\sigma$ lower & 1.06 & 0.94\\
\hline
\end{tabular}
\label{y500_tab}
\end{center}
\end{table}

\section{Conclusions}
There is an inconsistency between X-ray and SZE fitted
model parameters that leads to a bias in deduced values of $Y_{500}$ and
$M_{500,gas}$ when using isothermal $\beta$-models. Using our U-$\beta$
model reduces the bias and scatter, resulting in a more precise and
accurate determination of the $y$-$M$ scaling relation for
clusters.  

X-ray and SZE radial profiles from the same galaxy cluster can
not be fit well by identical isothermal $\beta$-models. The stronger
dependence of the SZE emission on temperature leads to problems when
the cluster temperature declines with radius. In contrast, the strong
dependence of the X-ray emission on density minimizes the error
introduced by variations in cluster temperature. We show that fitting
either the X-ray or SZE profiles to a modified, non-isothermal
$\beta$-model corrected by the inclusion of a univeral temperature
profile for clusters results in better consistency with the real
values of $Y_{500}$ and $M_{500,gas}$.

We expect this inconsistency should affect measurements of the Hubble
constant at some level, indeed it appears to result in a bias in
$H_0$, as shown by \citet{ameglio}. There will
certainly be an effect on gas fraction determinations, though it will
also depend on any bias inherent in X-ray hydrostatic total mass
estimates. While current precision in SZE/X-ray derived quantities is
not high enough to reveal effects at the 10-20\% level of the type
described here, we expect it will be higher with larger samples and new
instruments available in the near term. When contemplating clusters as
precision cosmological tools, effects as this level must be considered.

While some have used double $\beta$-models \citep{laroque} to fit cluster radial
profiles with some success, a model like the U-$\beta$ is preferred, since
it is physically motivated by the observed ICM properties. This work
illustrates a simple
modification of the isothermal $\beta$-model to more realistically
account for cluster physics, and thus more accurately and precisely
measure cluster properties.   
\acknowledgments
The simulations presented in this work were conducted at the
National Center for Supercomputing Applications at the University of
Illinois, Urbana-Champaign through computer allocation grant
AST010014N. We also acknowledge the support of the NSF through grant AST-0407368.
\appendix
\section{Derivation of UTP Surface Brightness Model}
To derive a surface brightness model resulting from the UTP modified
$\beta$-model, we take the standard $\beta$-model for density,
\begin{equation}
  n_{e}(r) = n_{e0}\left(1 + \left(\frac{r}{r_{c}}\right)^{2}\right)^{-3\beta/2},
\end{equation}
and the UTP model for temperature,
\begin{equation}
  T(r) = \langle T \rangle_{500}T_{0}\left(1 + \left(\frac{r}{\alpha r_{500}}\right)^{2}\right)^{-\delta},
\end{equation}
and substitute them into the integral of the X-ray surface brightness,
\begin{equation}
S_X(b) = \frac{1}{4\pi(1+z)^4}\int n_e(r) n_H(r) \Lambda(T(r)) dl.
\end{equation}
In the simple bremsstrahlung case, $\Lambda(T)$ = $\Lambda_0
T^{1/2}$. In that case,
\begin{equation} 
S_X(b) = S_{X0} \left( 1 + \left(\frac{b}{r_c} \right)^2
\right)^{-3\beta} \left(1+\left(\frac{b}{\alpha r_{500}} \right)^2
\right)^{-\frac{\delta}{2}} I_X(b),
\end{equation}
where 
\begin{equation}
S_{X0} = \frac{2{n_{e0}}^2 \Lambda_0(\langle T \rangle_{500}
  T_0)^{1/2}}{4\pi\kappa(1+z)^4},
\end{equation}
where $\kappa = \mu_H/\mu_e$, the ratio of the mean molecular weights of hydrogen and electrons. For $r_c < \alpha r_{500}$,
\begin{equation}
I_X(b) = \frac{r_c}{2} \left( 1 + \left(\frac{b}{r_c}\right)^2 \right)^{1/2}
B\left(\frac{1}{2},3\beta + \frac{\delta}{2} - \frac{1}{2} \right)
F_{2,1} \left(\frac{\delta}{2}, \frac{1}{2};3\beta +
\frac{\delta}{2},1-\frac{{r_c}^2 + b^2}{\alpha^2 {r_{500}}^2 + b^2}
  \right),
\end{equation}
where $F_{2,1}$ is Gauss' hypergeometric function and $B$ is the Beta
function defined by 
\begin{equation}
B(x,y) = \frac{\Gamma(x) \Gamma(y)}{\Gamma(x+y)};x,y > 0.
\end{equation}

 For $r_c > \alpha
r_{500}$, 
\begin{equation}
I_X(b) = \frac{\alpha r_{500}}{2} \left( 1 + \left(\frac{b}{\alpha r_{500}}\right)^2 \right)^{1/2}
B\left(\frac{1}{2},3\beta + \frac{\delta}{2} - \frac{1}{2} \right)
F_{2,1} \left(3\beta, \frac{1}{2};3\beta +
\frac{\delta}{2},1-\frac{\alpha^2 {r_{500}}^2 + b^2}{{r_c}^2 + b^2}
  \right).
\end{equation}

Similarly for the SZE, where 
\begin{equation}
y = \int \sigma_T n_e(r) \frac{k_b T}{m_e c^2} dl,
\end{equation}
the substitution results in 
\begin{equation}
y(b) = y_{0} \left( 1 + \left(\frac{b}{r_c}\right)^2
\right)^{-\frac{3\beta}{2}} \left(1+\left(\frac{b}{\alpha r_{500}} \right)^2
\right)^{-\delta} I_{SZ}(b),
\end{equation} 
and 
\begin{equation}
y_0 = 2\sigma_T n_{e0} \frac{k_b \langle T \rangle_{500} T_0}{m_e c^2}.
\end{equation}
For the two cases of $r_c < \alpha r_{500}$ and $r_c > \alpha
r_{500}$ respectively, 
\begin{equation}
I_{SZ}(b) = \frac{r_c}{2} \left( 1 + \left(\frac{b}{r_c}\right)^2 \right)^{1/2}
B\left(\frac{1}{2},\frac{3\beta}{2} + \delta - \frac{1}{2} \right)
F_{2,1} \left(\delta, \frac{1}{2};\frac{3\beta}{2} +
\delta,1-\frac{{r_c}^2 + b^2}{\alpha^2 {r_{500}}^2 + b^2}\right),
\end{equation}
and
\begin{equation}
I_{SZ}(b) = \frac{\alpha r_{500}}{2} \left( 1 + \left(\frac{b}{\alpha r_{500}}\right)^2 \right)^{1/2}
B\left(\frac{1}{2},\frac{3\beta}{2} + \delta - \frac{1}{2} \right)
F_{2,1} \left(\frac{3\beta}{2}, \frac{1}{2};\frac{3\beta}{2} +
\delta,1-\frac{\alpha^2 {r_{500}}^2 + b^2}{{r_c}^2 + b^2}
  \right).
\end{equation}

\begin{figure}
\begin{center}
\includegraphics[width=0.6\textwidth]{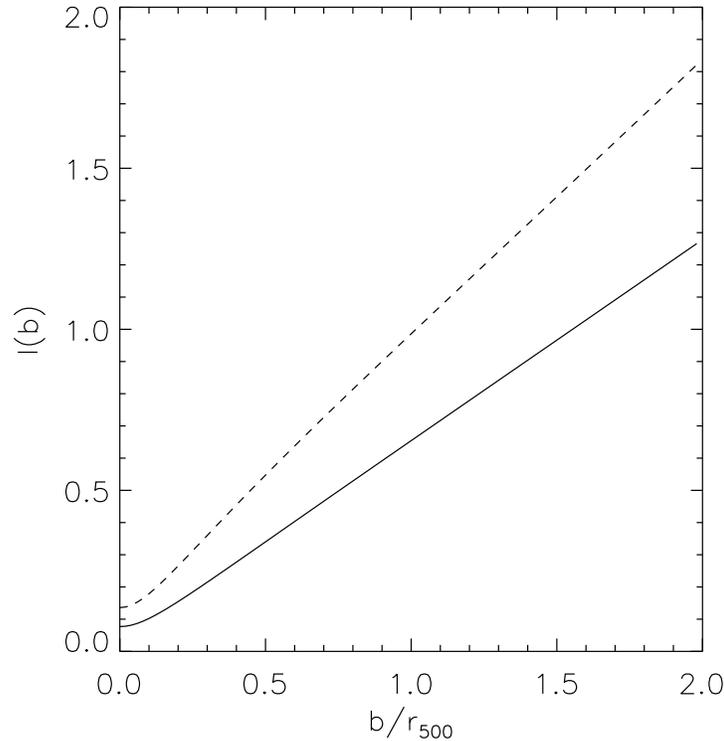}
\end{center}
\caption{Value of line integrals $I_X$ and $I_{SZ}$as a function of radius normalized to
  $r_{500}$. This calculation uses values near the mean of the
  distribution for all clusters for the $\beta$-model paramters, and
  $r_{500}$=1.0$h^{-1}$Mpc. $r_{core}$=160kpc, $\beta$=0.82,
  $\alpha$=1.15, $\delta$=0.5, therefore this is the case where
  $r_{core} < r_{500}$. Solid line is for X-ray integral,
  dotted is for SZE.}
\label{ival}
\end{figure}

Figure \ref{ival} shows the values for each of the line integrals as a function
of projected radius normalized to $r_{500}$ for typical model
parameters. We have created a calculator for the line integral in the
Interactive Data Language (IDL), which is available on-line at http://solo.colorado.edu/~hallman/UTP/Ib\_calc.tar.gz.

\end{document}